# Execution and Result Integration Scheme in FPU Farms for Co-ordinated Performance

[1] T.R.Gopalakrishnan Nair, , [2]R. Selva rani, [3]Vighnaraju Saraf
[1]Director, Research & Industry Incubation Center, Senior Member IEEE, trgnair@ieee.org
[2]Associate Professor, Research-CSE, selvss@yahoo.com
[3]M.Tech II Year, s.vighnaraju@gmail.com
[1,2,3] Dayananda Sagar Institutions, Bangalore, India

**Abstract - The main goal of this research is to develop the concept of an innovative processor system called Functional Processor System. The particular work carried out in this paper focuses on the execution of functions in the heterogeneous functional processor units(FPU) and integration of functions to bring net results. As the functional programs are super-level programs, the requirements of execution are only at functional level. The Execution and integration of results of functions in FPUs are a challenge. The methodology of executing the functions in the functional processor farm and the integration of results of functions according to the assigned addresses are investigated here. The concept of feeding the functions into the processor is promoted rather than the processor fetching the instructions/functions and executing in this paradigm. This work is carried out at conceptual levels and it takes a long way to go into the realization of this model in hardware, possibly only with a large industry team and with a realistic time frame.**

## 1. INTRODUCTION

Computer architecture forms the bridge between the application needs and the capabilities of the underlying technologies. As application demands change and technologies cross various thresholds computer architects must continue the innovation to generate systems that can deliver improved performance and cost effectiveness. To design a leadership format of computer systems, we must thoroughly understand the nature of the workloads that such systems are intended to support.

Even as demands of applications for computational power continue to grow, silicon technology is running into some major discontinuities as it scales to challenging smaller sizes. From a study on operating frequencies of microprocessors it is seen that the operating frequencies of microprocessors introduced over the last 10 years and projected frequencies for the next two to three years, will grow in the future at half the rate of the past decade. We need 80-plus percent compound growth in system-level performance, while frequency growth has dropped to 15- 20% because of power limitations [1]. The computer architecture community's challenge therefore, is to devise innovative ways of delivering continuing growth in system performance and price-performance while simultaneously solving the space and power problem over the silicon. It must lead to a manageable and scalable architecture not having the bottlenecks of the design trend of 20$^{th}$ century. Rather than riding only on the steady frequency intensification and duplication of foot print of units of the past decade, system performance improvements will increasingly be driven by integration at all levels, together with hardware-software optimization.

Although most current multicore processors are homogeneous, micro architects are now proposing heterogeneous core implementations, including systems in which heterogeneity is introduced at runtime. The primary problem with homogeneous multicore processors is that, naive replication of state of the art single-core designs in a single package (or chip package), stresses the power and cooling limits for the chip.[2],[4] .

To overcome the limitations of general purpose microprocessors and incapability due to conventionalism of programming environment, an innovative architecture called Functional Processors, have been developed. It is a heterogeneous function processor array comprising of multiple cores, where in, each core is specialized in performing functions of a selected domain having connotations to regional learning approach of human brain.

The main goal of this research is to generate a viable functional processor system whereby a program is represented as a sequence of higher level functions only and executed on multiple functional processor units simultaneously. It also intents to demonstrate a deviation from conventional instruction fetching system to a function feed system to the executing nodes.

## 2. FUNCTION PROCESSOR ARCHITECTURE

Figure.1 shows theoverview of the Functional Processor Architecture. The Functional processor Architecture (FPA) has eight high-frequency specialized execution cores with pipelined MIMD capabilities and aggressive function transfer architecture. Functional processor architecture - is an innovative solution whose design is based on the analysis of a broad range of workloads in areas such as graphics applications and lighting, physics, fast-Fourier transforms (FFT), matrix operations, cryptography, scientific workloads, general business patterns as functions and in future self learning of functions at nodes. Functional programs generated with a functional chain generation approach for any application. Functional decompositions can be achieved prior to execution by splitting functions from the application. Functional decompositions can be static or dynamic, and they should be orchestrated carefully to fully utilize the FPUs.

The FPA provides programming support for using some of the aforementioned concurrent execution strategies, selecting the most effective strategy. Actually, combining and scheduling layered parallelism on FPA can be an arduous task. To simplify programming and improve efficiency on FPUs we use the feed mechanism rather than using fetching mechanism.

The FPA defines five separate types of Functional components: the Functional Decoder, Fine Decoding, Functional Processor Unit (FPU), Functional Processor Interconnect bus and the Local store (LS). Each FPU has a dedicated local storage and dedicated cache. The combination of these components is called a Functional Processor Unit or FPU.

An application contains an array of functions and the functions are pushed into the functional decoder. The functional decoder recognizes the functions from the process or the application. The functional decoder isolates the functions say $Fn_1$, Fn2....$Fn_m$ and is stored in separate modules. These functions are shoved into the fine decoding block. Here in fine decoding block we have "Funpiler", which is the function compiler. It assigns the addresses to the functions, which are to be executed in the corresponding FPU farm. The Funpiler analyses the functions say, if the function desires graphical processor then the "Function ID"-FID G1, FID G2 addresses are assigned to the functions. If it desires arithmetic processor then FID-A1 is assigned. If the FID of the function, maps with the FPU then the fine decoder assigns the function to it. The FPU executes the function. We explore the design and implementation of our scheduler using Lamport's bakery algorithm. It maintains the first-come-first serve property by using a distributed version of the number dispensing machines often found in bakeries. Each function takes a number in the doorway, and then waits until no function with an earlier number is trying to enter it.

The FIFO scheduling is implemented in fine decoding block. The function rules the FPU till its execution. The main challenge here is the alignment of the functions in the integration unit. The functions are aligned according to the assigned addresses and it is stored in the memory for further use or else displayed in the display unit. The alignment of the functions plays a significant role.

## 3. FUNCTION STATES

During the execution of functions, the functions change its state. The state of the process is the current condition or status of the process. In a POSIX –model environment, a process can be in the following states:

- Running
- Runnable (ready)
- Waiting (blocked)
- Stopped

The current condition of the function depends upon the circumstances created by the process or by the operating system. When certain circumstances exist, the process will change its state. State transition is the circumstance that causes the function to change its state. Figure.2 is the state diagram for the function states. The state diagram has nodes and directed edges between the nodes. Each node represents the state of the process. The directed edges between the nodes are state transitions. Table.2 lists the state transitions with a brief description. Figure 2 and Table 2 show, only certain transitions are allowed between states. For example, there is a transition, an edge, between ready and running, but there is no transition, no edge, between sleeping and running. Meaning, there are circumstances that cause a process to move from the ready state to the running state, but there are no circumstances that cause a process to move from the sleeping state to a running state.

The actual priority of the function is based on its programmed priority minus a value that indicates how recently the function has actually run. This value is subject to continual adjustment. The more time passes, the closer to zero the value becomes. This primarily distinguishes between functions of the same priority, and it leads to round robin scheduling between functions of the same priority. All things being equal, each function of the same priority will receive approximately the same amount of FPU time.

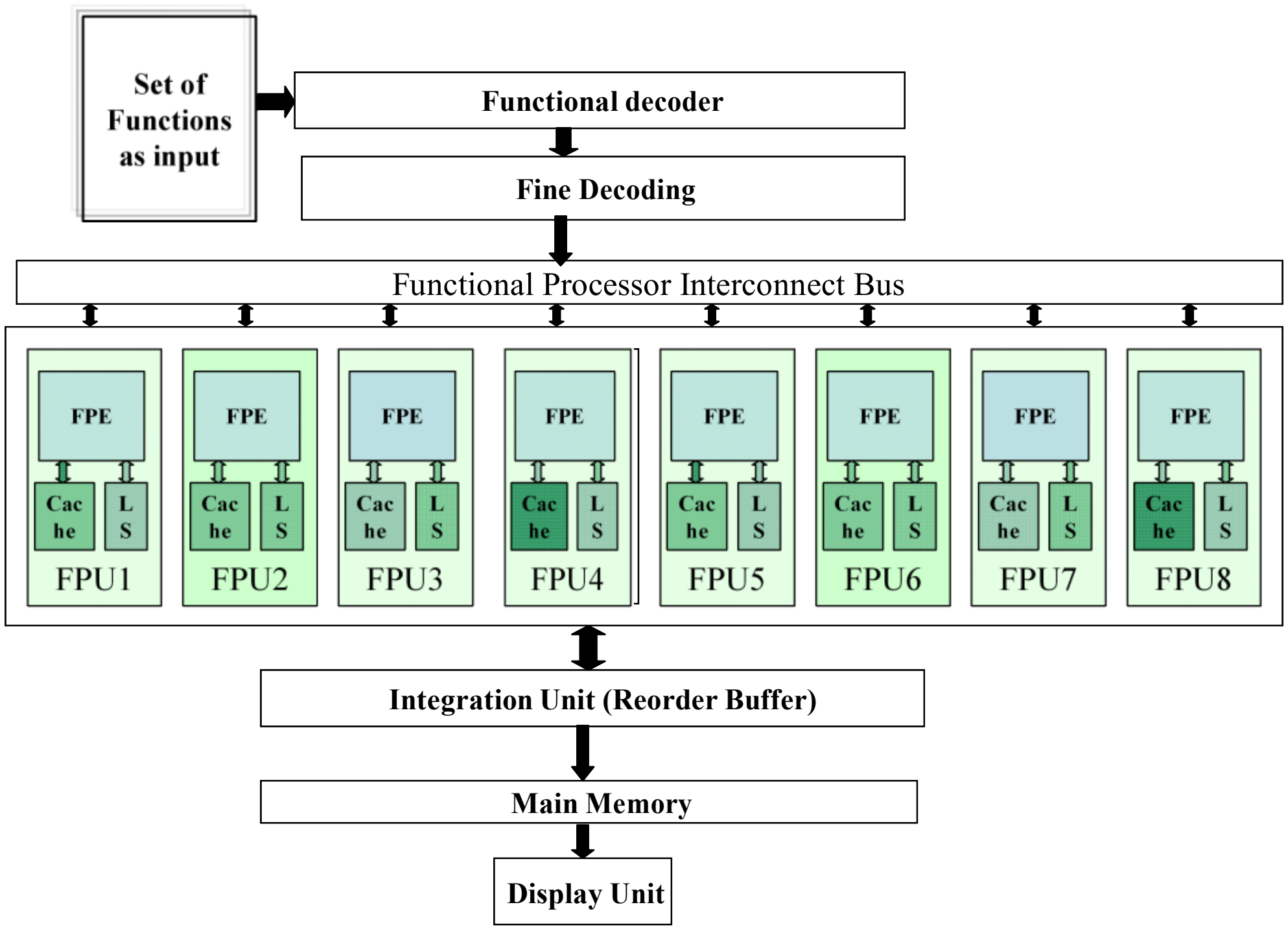

Figure 1. Functional Processor Architecture
FPU: Functional Processor Unit, FPE: Functional Processor Element LS: Local store

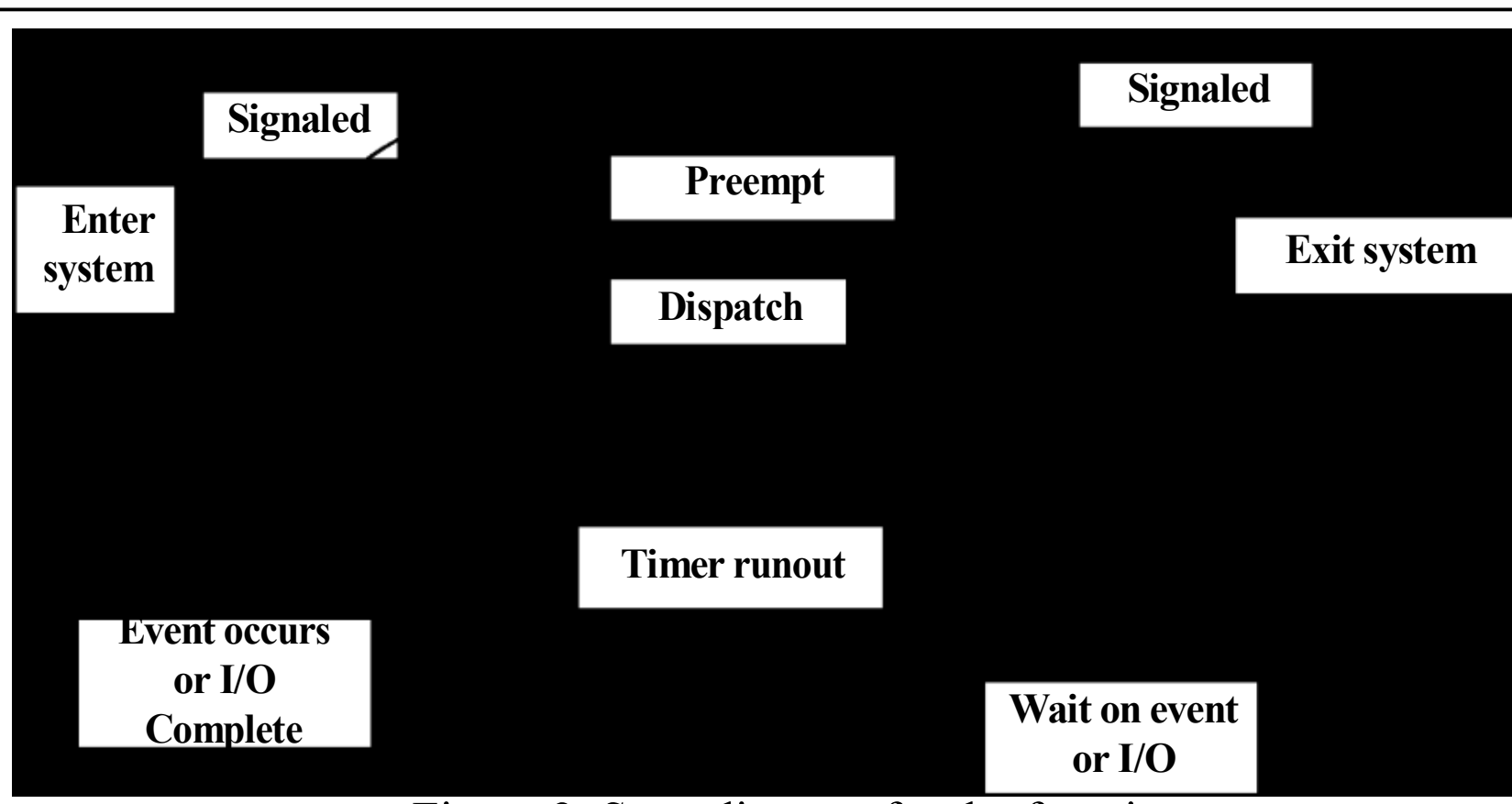

Figure 2. State diagram for the functions

Table 1 State Transition Table

| State Transitions | Descriptions |
|---|---|
| READY→RUNNING (dispatch) | The Function is assigned to the processor. |
| RUNNING→ READY(timer runout) | The time slice the function assigned to the processor has run out. The function is placed back in the queue. |
| RUNNING→ READY(preempt) | The function has been preempted before the time slice ran out. This can occur if a function with a higher priority is runnable. The function is placed back in the ready queue. |
| RUNNING→ SLEEPING (block) | The function gives up the processor before the time slice has run out. The function may need to wait for an event or has made a system call, for example, a request for I/O. The function is placed in a queue with other sleeping functions. |
| SLEEPING→READY (unblock) | The event the function was waiting for has occurred, or the system call has completed. For example, the I/O request is filled. The process is placed back in the ready queue. |
| RUNNING→ STOPPED | The function gives up the processor because it has received a signal to stop. |
| STOPPED→READY | The function has received the signal to continue and is placed back in the ready queue. |
| RUNNING→EXIT | The function has terminated, the parent has retrieved the exit status, and the function |

## 4. COMMUNICATION AND SYNCHRONIZATION OF CONCURRENT TASKS

If communication between dependent functions is not appropriately designed, then data race conditions can occur. Determining the proper co-ordination of communication and synchronization between functions requires matching the appropriate concurrency models during problem and solution decomposition. Concurrency models dictate how and when communication occurs and the manner in which work is executed.

### 4.1 Dependency Relationships

When a function requires communication or co-operation among each other to accomplish a common goal, they have a dependency relationship. Fn1 depends on Fn2 to supply value for a calculation, to give the name of the file to be processed, or to release a resource. Fn1 may depend on Fn2, but Fn2 may not have a dependency on Fn1. Given any two tasks, there are exactly four dependency relationships that can exist between them:

Table 2: Dependency relationships

| | |
|---|---|
| A→B | Fn1 depends on Fn2. |
| A ← B | Fn2 depends on Fn1 |
| A ↔ B | Fn1 depends on Fn2, and Fn2 depends on Fn1 |
| A NULL B | There are no dependencies between Fn1 and Fn2. |

In the first and second cases, the dependency is a one way unidirectional dependency. In the third case, there is a two way bi-directional dependency; Fn1 and Fn2 are mutually dependent on each other. In the fourth case, there is a NULL dependency between Fn1 and Fn1; no dependency exists.

### 4.2 Communication Dependencies

Functions can communicate with other functions within the address space of their process by using global variables and data structures. If two functions wanted to pass data between them, $Fn_1$ would write the name of the file to a global variable, and $Fn_2$ would simply read that variable. These are examples of unidirectional communication dependencies where only one task depends on another task.

### 4.3 Counting Function Dependencies

The overall task relationships between the functions in an application by enumerating the number of possible dependencies that exist. The possible dependencies and then their relationships determine which function must be coded for communication and synchronization. This is similar to truth tables used to determine possible branches of decision in a program or application. For example, if there are three functions A, B, and C, the possible communication dependencies that exist among the functions. If there are two functions involved in a dependency, use combination to calculate the possible functions involved in the dependency from the three functions:

$$C\ (n,\ k)$$

Where,
***n*** is the number of functions
***k*** is the number of functions involved in the dependency.

So, for the example *C (3, 2)*, the answer is 3; there are three possible combinations of functions: A and B, A and C, B and C. Now if you consider each combination as a graph (with two nodes and one edge between them), a simple graph, meaning that there are no self - loops and no parallel edges (no two edges will have the same endpoints), then the number of edges in a graph is $n\ (n - 1)/2$. So, for the two - node simple graph, there are $2(2 - 1)/2$, which is 1. There is one edge for each graph. Now each edge can have four possible dependency relationships as discussed in Table II.

So, each individual graph has four possible relationships. The number of possible dependency relationships among three functions in which two are involved in the relationship. There are 12 possible relationships. An adjacency matrix can be used to enumerate the actual dependency relationships for two – function combinations. An adjacency matrix is a graph.

$$G = (V, E)$$

Where,
*V* is the set of vertices or nodes of the graph and *E* is the set of edges such that:

$A(i, j) = 1$ if $(i, j)$ is an element of $E$
$= 0$ otherwise
$A(i, j) <> A(j, i)$

Where ***i*** denotes a row and ***j*** denotes a column. The size of the matrix is ***n* x *n*,** where ***n*** is the total number of functions. Figure. A shows the adjacency matrix for three functions. The 0 indicates that there is no dependency, and the 1 indicates that there is a dependency. An adjacency matrix can be used to demarcate all of the dependency relationships between any two functions. On a diagonal, there are all 0s because there are no self - dependencies.

***A(1,2) = 1*** means for A → B, A depends on B.

***A(1,3) = 0*** means for A → C, A does not depend on C.

***A(2,1) = 0*** means for B → A, B does not depend on A.

***A(2,3) = 1*** means for B → C, B depends on C.

***A(3,1) = 1*** means for C → A, C depends on A.

***A(3,2) = 0*** means for C → B, C does not depends on B.

| | A | B | C |
|---|---|---|---|
| A | 0 | 1 | 0 |
| B | 0 | 0 | 1 |
| C | 1 | 0 | 0 |

Figure 1. Adjacency Matrix

| | A | B | C |
|---|---|---|---|
| A | | S,C0 | |
| B | | | S,C |
| C | S,C | | |

Figure 2. Dependency Matrix

A dependency graph is useful for documenting the type of dependency relationship, for example, *C* for communication or *Co* for cooperation. *S* is for synchronization if the communication or cooperation dependency requires synchronization. To construct the dependency graph, the adjacency matrix is used. Where there is a 1 in a row column position, it is replaced by the type of relationship. Figure (B) shows the dependency graph for the three functions. The 0s and 1s have been replaced by *C* or Co. Where there was a 0, no relationship exists; the space is left blank. For *A(1,2)*, A depends on B for synchronized cooperation, *A(2,3)* B depends on C for synchronized communication, and *A(3,2)* C depends on A for synchronized communication. Bidirectional relationships like A ↔ B can also be represented, but there are none in this example. So, all of the relationships can be represented in the matrix. For a NULL relationship a 0 is used in the adjacency matrix, and in the dependency matrix that position will be left blank.

## 5. COMMUNICATION BETWEEN FUNCTION PROCESSING ENTITIES

Functions also do not require special mechanisms for communication with other functions of the process called peer functions. Functions can directly pass and receive data from other peer functions. This saves system resources that would have to be used in the setup and maintenance of special communication mechanisms if multiple processes were used. Functions communicate by using the memory shared within the address space of the process. For example, if a queue is globally declared by a process function Fn1 of the process can store the name of a file that peer function Fn3 is to process. Fn3 can read the name from the queue and process the data. Processes can also communicate by shared memory, but processes have separate address spaces and, therefore, the shared memory exists outside the address space of both processes.

If you have a process that also wants to communicate the names of files it has processed to other processes, you can use a message queue. It is set up outside the address space of the processes involved and generally requires a lot of setup to work properly. This increases the time and space used to maintain and access the shared memory.

We have employed a graph theoretical approach to analyze the function execution in the function processing units. In the Figure.3, the function processing entities are represented by the notation C1, C2 up to C8. The functions are represented using the notations $F_1$, F2 upto $F_{21}$.

## 6. SYNCHRONIZING CONCURRENCY AND INTEGRATION

In any computer system, the resources are limited. There are limitations on memory, I/O devices and ports, hardware interrupts for processors cores to go around. The number of I/O devices is usually restricted by the number of I/O ports and the hardware interrupts that a system has. In an environment of limited hardware resources, an application consisting of multiple processes and functions must compete for memory locations, peripheral devices and processor time. Some functions and processes will be working together intimately using the system's limited sharable resources to perform a task and achieve a goal while other functions and processes work asynchronously and independently competing for those same sharable resources.

It is the operating system's job to determine when the process or function utilizes system resources and for how long. With preemptive scheduling, the operating system can interrupt the process or function in order to accommodate all the processes and functions competing for the system resources. There are software resources and hardware resources. An example of software resources is a shared library that provides a common set of services or functions to processes.

In the Integration block, all the functions are integrated according to the assigned addresses and finally the results are either stored in the memory or displayed at the output unit.

## BIOGRAPHY

**T.R. Gopalakrishnan Nair** holds M.Tech. (IISc, Bangalore) and Ph.D. degree in Computer Science. He has 3 decades experience in Computer Science and Engineering through research, industry and education. He has published several papers and holds patents in multi domains. He has won the PARAM Award for technology innovation. Currently he is the Director of Research and Industry in Dayananda Sagar Institutions, Bangalore, India.

**R. Selvarani** holds MI.Tech., Ph.D. in Computer Science and Engineering (Thesis submitted) and has 18 years of experience in teaching and research. She has published several research papers in computer science and holds two patents. She has been awarded Best Teacher Award twice in various institutions. Currently she is working as a Professor, Research (CSE) in Research and Industry incubation centre, Dayananda Sagar Institutions, Bangalore, India.

**Vighnaraju Saraf** holds M.Tech degree (Thesis submitted) in Digital Communication and Networking branch. He served as Lecturer for 2 years. He is a Fellow of IEEE**.**